\documentclass[twocolumn,showpacs,preprintnumbers]{revtex4}

\input{epsf}

\def\GRG{{\it Gen. Relativity and Gravitation} }

\def\PL{{\it Phys. Lett.} }
\def\PR{{\it Phys. Rev.} }
\def\PRL{{\it Phys. Rev. Lett.} }

\def\PTP{{\it Progr. Theor. Phys.} }

\def\al{\alpha}
\def\be{\beta}
\def\ga{\gamma}
\def\de{\delta}

\def\th{\theta}

\def\ka{\kappa}
\def\la{\lambda}

\def\Ga{\Gamma}
\def\De{\Delta}

\def\Om{\Omega}
\def\mn{{\mu\nu}}

 \def\frac#1#2{{\textstyle{{#1}\over
{#2}}}} 
\def\lsim{\mathrel{\rlap{\lower4pt\hbox{\hskip1pt$\sim$}}
    \raise1pt\hbox{$<$}}} \def\gsim{\mathrel{\rlap{\lower4pt\hbox{\hskip1pt$\sim$}}
    \raise1pt\hbox{$>$}}}
\def\sqr#1#2{{\vcenter{\vbox{\hrule height.#2pt
         \hbox{\vrule width.#2pt height#1pt \kern#1pt
         \vrule width.#2pt}
         \hrule height.#2pt}}}} 

\def\beq{\begin{equation}}
\def\eeq{\end{equation}}
\def\beqa{\begin{eqnarray}}
\def\eeqa{\end{eqnarray}}

\begin{document}

\title{The flight of the bumblebee:
\par vacuum solutions of a gravity model with vector-induced \par spontaneous Lorentz
symmetry breaking}

\vskip 0.2cm

\author{O. Bertolami, J. P\'aramos}

\vskip 0.2cm

\affiliation{Instituto Superior T\'ecnico, Departamento de
F\'{\i}sica, \\ Av. Rovisco Pais 1, 1049-001 Lisboa, Portugal}

\vskip 0.2cm

\affiliation{E-mail addresses: orfeu@cosmos.ist.utl.pt;
x\_jorge@netcabo.pt}

\vskip 0.5cm

\date{\today}

\begin{abstract}
We study the vacuum solutions of a gravity model where Lorentz
symmetry is spontaneously broken once a vector field acquires a
vacuum expectation value. Results are presented for the purely
radial Lorentz symmetry breaking (LSB), radial/temporal LSB and
axial/temporal LSB. The purely radial LSB result corresponds to
new black hole solutions. When possible, Parametrized
Post-Newtonian (PPN) parameters are computed and observational
boundaries used to constrain the Lorentz symmetry breaking scale.

 \vskip 0.5cm

\end{abstract}

\pacs{04.25.Nx, 11.30.Cp, 11.30.Qc \hspace{2cm} Preprint
DF/IST-3.2005}

\maketitle

\section{Introduction}

Lorentz invariance is possibly one of the most fundamental
symmetries of Nature. It is theoretically sound, it has been
extensively tested, and so far no clear cut experimental evidence
has emerged to bridge its validity (see e.g. Refs.
\cite{Kostelecky1, Bertolami}). It is therefore no surprise that
most theories of gravity encompass this symmetry, and little
attention has been paid in understanding the implications of the
breaking of Lorentz invariance in this context.

There is, however, a window of opportunity for fiddling with a
less stringent approach: as for gauge symmetries in field theory,
many relevant after effects can arise if one assumes a spontaneous
breaking of Lorentz invariance \cite{Kostelecky2, Kostelecky3}. In
the context of gravity, the breaking of Lorentz invariance can be
implemented if a vector field ruled by a potential exhibiting a
minimum rolls to its vacuum expectation value (\textit{vev}), in
the fashion of the Higgs mechanism \cite{Kostelecky4}. This
``bumblebee'' vector thus acquires an explicit (four-dimensional)
orientation, and preferred-frame effects may arise.

The possibility of violation of the this fundamental symmetry has
been widely discussed in the recent literature (see Refs.
\cite{Kostelecky1, Bertolami}). Indeed, string/Mtheory allow for a
spontaneous breaking of Lorentz symmetry, due to the existence of
non-trivial solutions in string field theory \cite{Kostelecky2,
Kostelecky3}, loop quantum gravity \cite{LSB1, LSB2}, quantum
gravity inspired spacetime foam scenarios \cite{LSB3} and
noncommutative field theories \cite{LSB4, LSB5}. Also, LSB could
result from spacetime variation of fundamental coupling constants
\cite{LSB6}. Experimental tests of this symmetry breaking could be
achieved, for instance, in ultra-high energy cosmic rays
\cite{LSB7}.

Efforts to quantify an hypothetical breaking of Lorentz invariance
have been mainly concerned with the phenomenology of such
spontaneous Lorentz symmetry breaking (LSB) in particle physics.
Only recently its implications for gravity have been more
thoroughly explored \cite{Kostelecky4, Kostelecky5}. In that work,
the framework for the LSB gravity model is set up, developing the
action and using the \textit{vielbein} formalism.

In this study, we focus on the consequences of such scenario. For
this, one assumes three relevant cases: the bumblebee field
acquiring a purely radial \textit{vev}, a mixed radial and
temporal \textit{vev} and a mixed axial and temporal \textit{vev}.
For prompt comparison with experimental tests, we shall write our
results in terms of the PPN parameters, when possible.

The action of the bumblebee model is written as

\beqa \nonumber S & = & \int d^4 [ {\sqrt{-g} \over 2 \ka} \left(
R + \xi B^\mu B^\nu R_{\mu\nu} \right) \\ & - &  {1 \over 4}
B^{\mu\nu} B_{\mu\nu} - V(B^\mu B_\mu \pm b^2 ) ]~~, \eeqa

\noindent where $\ka = 8 \pi G$, $\xi$ is a real coupling constant
and $b^2$ is a real positive constant. The potential $V$ driving
Lorentz and/or CPT violation can be chosen to have a minimum at $
B^\mu B_\mu \pm b^2 = 0 $.

\section{Purely radial LSB}

In this section we develop a method to obtain the exact solution
for the purely radial LSB. For this, we assume a static,
spherically symmetric spacetime, with a metric $ g_\mn =
diag(-e^{2\phi},e^{2 \rho},r^2,r^2 sin^2\th)$, where $\phi$ and
$\rho$ are functions of $r$. Also, we admit that the vector field
$B_\mu$ has relaxed onto its expectation value $b_\mu$. It is
imposed that $D_\mu b_\nu = 0$. This enables us to calculate
$b_\mu$, using the affine connection derived from the metric
$g_\mn$; for this, and since the only non trivial covariant
derivative is with respect to the radial coordinate, it is assumed
that $b_\mu = (0, b(r), 0, 0)$. Hence, from

\beq D_\mu b_\nu = \partial_\mu b_\nu - \Ga^\al_{\nu\mu} b_\al = 0
~~, \eeq

\noindent follows that

\beq b(r) = \xi^{-1/2} b_0 e^{\rho}~~, \eeq

\noindent where the $ \sqrt{\xi}$ is introduced for later
convenience. It can be immediately understood that, as expected, $
b^2 = b^\mu b_\mu = b_0^2 \xi^{-1} $ is constant. The action can
be thus written as

\beq S_s = \int d^4 x \sqrt{-g} \left[ {R \over 2 \ka} +
(g^{rr})^2 b^2(r) R_{rr} \right]_s ~~, \eeq

\noindent where the subscript $\textit{s}$ stands for the spatial
quantities. With the assumed metric, the determinant is given by
$\sqrt{-g} = r^2 e^{\rho + \phi}$; the scalar curvature and the
relevant Ricci tensor non-vanishing component are given by

\beq R = { 2 \left[1 + (2 r \rho' -1) e^{-2\rho} \right] \over
r^2}~~,~~ R_{rr} = {2 \rho' \over r} ~~, \eeq

\noindent where the prime stands for derivative with respect to
$r$ and we have integrated over the angular dependence. Also,

\beq \xi (b^r)^2 R_{rr} = b_0^2 {2 \rho' e^{-2\rho} \over r}~~,
\eeq

\noindent where $b^r$ is the (contravariant) radial component of
$b_\mu$. We now introduce the field redefinition $ \Psi =
\left(1-e^{-2\rho} \right) r^{-2}$, so that

\beq \Psi' = {2 \rho' e^{-2 \rho} \over r^2} - {2 \Psi \over r}~~,
\eeq

\noindent and thus

\beqa \nonumber {2 \rho' e^{-2\rho} \over r} & = & 2 \Psi + r
\Psi' ~~, \\ \nonumber R & = & 2 (3 \Psi + r \Psi') ~~, \\
(b^r)^2 R_{rr} & = & b_0^2 (2\Psi + r \Psi') ~~. \eeqa

We can now work with the action (see \cite{Bento} and references
therein):

\beq S_s = {2 \over \ka} \int dr e^{\rho + \phi} r^2 \left[ ( 3 +
b_0^2 ) \Psi + ( 1 + {b_0^2 \over 2} ) r \Psi' \right]~~.
\label{spaction} \eeq

Variation with respect to $\phi$ yields the equation of motion

\beq ( 3 + b_0^2 ) \Psi + ( + 1 + {b_0^2 \over 2} ) r \Psi' = 0
~~, \eeq

\noindent which admits the solution $ \Psi(r) = \Psi_0 r^{-3+L}$,
where we define

\beq 3 - L \equiv {3 + b_0^2 \over 1 + {b_0^2 \over 2}} \simeq 3 -
{b_0^2 \over 2}~~, \eeq

\noindent and hence $L \simeq b_0^2 /2$. From the definition of
$\Psi$, we obtain

\beq g_{rr} = e^{2 \rho} = \left( 1 - \Psi_0 r^{-1+L}
\right)^{-1}~~. \eeq

Comparing with the usual Schwarzschild metric, one can write

\beqa \nonumber g_{rr} & = & \left( 1 - {2 G_L m \over r} r^L \right)^{-1}~~, \\
g_{00} & = & -1 + {2 G_L m \over r} r^L~~, \label{metric} \eeqa

\noindent where $G_L$ has dimensions $ [ G_L ] = L^{2-L} $ (in
natural units, where $c=\hbar=1$). One can define $G_L = G
r_0^{-L}$, where $r_0$ is an arbitrary distance. The $L
\rightarrow 0$ limit yields $G_L \rightarrow G$ and the usual
geometrical mass, $M \equiv G m $, with dimensions of lenght. From
now on we express all results in terms of $M$. The event horizon
condition is given by

\beq g_{00} = -1 + {2 M \over r} {r \over r_0}^L = 0 ~~, \eeq

\noindent thus $ r_s = (2 M r_0^{-L} )^{1/1-L} $. One can compute
the scalar invariant $I = R_{\mu \nu \rho \la}R^{\mu \nu \rho
\la}$ (the norm-square of the Riemann tensor), obtaining

\beqa \nonumber I & = & 48 \left[ 1 - {5 \over 3} L + {17 \over
12} L^2 - {1 \over 2} L^3 + {1 \over 12} L^4 \right] M^2 \left({r
\over r_0}\right)^{2L} r^{-6} \\ & \simeq & \left( 1 - {5 \over 3}
L\right)  \left({r \over r_0}\right)^{2 L} I_0 ~~, \eeqa

\noindent where $I_0 = 48 M^2 r^{-6}$ is the usual scalar
invariant in the limit $L \rightarrow 0$. Hence, since $I(r=r_s)$
is finite, one sees that the singularity at $r = r_s$ is
removable. By the same token, the singularity at $r=0$ is
intrinsic, as given by the divergence of the scalar invariant.
Thus, an axial LSB gravity model admits new black hole solutions
whose singularity is well protected within a horizon of radius
$r_s$. One expects a Hawking temperature given by

\beq T = {\hbar \over k_B} {1 \over 4 \pi r_s } = (2M
r_0^{-L})^{-L/(1-L)} T_0 \simeq (2M r_0^{-L})^{-L} T_0 ~~, \eeq

\noindent where $T_0 = \hbar / 8 \pi k_B M $ is the usual Hawking
temperature, which is recovered in the limit $L \rightarrow 0$.

Of course, this exact solution does not allow for a PPN expansion,
as the obtained metric cannot be expanded in powers of $U=M/r$.
One can, however, compare with results for deviations from
Newtonian gravity \cite{Fischbach}, usually stated in terms of a
Yukawa potential of the form

\beq V_Y(r) = {G_Y m \over r} \left( 1 + \al e^{-r/\la} \right)~~.
\eeq

\noindent Equating this potential to the one arising from $g_{00}$
in Eq. (\ref{metric}), one gets

\beq G_L r^L = G_Y \left(1 + \al e^{-r/\la} \right)~~. \eeq

\noindent Expanding to first order around $r = r_0$ yields

\beq G_L r_0^L \left(1 + L {r \over r_0} \right) = G_Y \left(1 +
\al - \al {r \over \la} \right)~~, \eeq

\noindent which allows us to identify $\la = r_0$ and $\al = -L$
(with $G_Y (1-L) = G_L r_0^L = G$). This states that the effects
of a radial LSB, probed at a distance $r_0$ from the source, can
be interpreted as due to a Yukawa potential of coupling strength
$\al = -L$ and range $\la = r_0$. The negative sign of the
coupling shows that the radial LSB yields a ``repulsive''
component (for $r > r_0$), as can be seen from Eq. (\ref{metric}).

Notice that this identification of the LSB effect with a Yukawa
potential constraints the length $\la$ to be equal to the distance
$r_0$ at which one tests deviations from the inverse square law.
This is not the case with a ``true'' Yukawa perturbation, where
each test of gravity at a distance $r_0$ yields an allowed range
for both $\al$ and $\la$ (although, of course, the test is
sensible to deviations only at a scale $\la$ close to $r_0$.
Hence, to obtain a bound on $L$ one must look at the allowed value
of $\al$ for the fixed $\la=r_0$ at which an experiment is carried
out. The most stringent bound is derived from planetary tests to
Kepler's law, with Venus yielding $\la = r_0 = 0.723~AU$ and $L =
|\al| \leq 2 \times 10^{-9}$ \cite{Fischbach}.

\section{Radial/temporal LSB}

We consider now the mixed radial and temporal Lorentz symmetry
breaking. As before, we assume that the bumblebee field $B_\mu$
has relaxed to its vacuum expectation value. Assuming temporal
variations to be of order of the age of the Universe $H_0^{-1}$,
where $H_0$ is the Hubble constant, one can, as before, consider a
Birkhoff static, radially symmetric metric $ g_\mn =
diag(-e^{2\phi},e^{2 \rho},r^2,r^2 sin^2(\th))$. Imposing as
(physical) gauge choice the vanishing of the covariant derivative
of the field $B_\mu$, one gets $b_r(r) = \xi^{-1/2} A_r e^{\rho}$
and, similarly, $ b_0(r) = \xi^{-1/2} A_0 e^{\phi}$, with $A_0$
and $A_r$ dimensionless constants. One immediately sees that, as
expected, $ b^2 = b^\mu b_\mu = (A_r^2 - A_0^2)\xi^{-1}$ is
constant.

Since one now has both a radial and a time component for the
vector field \textit{vev}, the symmetry $\phi = -\rho$ does not
hold. Therefore, one cannot use the previous spatial action
formalism depicted in Eq. (\ref{spaction}). Instead, one resorts
to the full Einstein equations,

\beq G_{\mu\nu} = \xi \left[ {1 \over 2} (b^\al)^2 R_{\al\al}
g_{\mu\nu} - b_\mu b^\nu R_{\mu\nu} - b_\nu b^\mu R_{\nu\mu}
\right] ~~.\eeq

The additional equation of motion for the vector field vanishes,
since it has relaxed to its \textit{vev} and therefore both the
field strength and the potential term are null. Introducing the
metric Ansatz and the expressions for $b_\mu$ one gets, after a
little algebra,

\beqa \nonumber G_{00} & = & {1\over 2} \left[3 A_0^2 R_{00} -
A_r^2 e^{2(\phi -\rho)} R_{rr} \right] ~~, \\ G_{rr} & = & {1\over
2} \left[A_0^2 e^{2 (\rho -\phi)} R_{00} - 3 A_r^2 R_{rr} \right]
~~,\eeqa

\noindent We now write $G_{00} = g_0(r) e^{2(\phi-\rho)}$, $G_{rr}
= g_r(r)$, $R_{00} = f_0(r) e^{2(\phi -\rho)}$ and $R_{rr} =
f_r(r)$, where

\beqa \nonumber f_0(r) & \equiv & {(2 - r \rho')\phi' \over r} + \phi'^2 + \phi''~~, \\
\nonumber f_r(r) & \equiv & {(2+r\phi')\rho' \over r} - \phi'^2 - \phi''~~, \\
\nonumber g_0(r) & \equiv & {-1 + e^{2 \rho} \over r^2} + {2\rho' \over r}~~, \\
g_r(r) & \equiv & {1 - e^{2 \rho} \over r^2} + {2\phi' \over r}~~.
\eeqa

\noindent Inserting the above into the Einstein equations, it
follows that

\beqa \nonumber \label{coupled} g_0(r) & = & {1\over 2} \left[3
A_0^2 f_0(r) - A_r^2 f_r(r) \right]~~, \\ g_r(r) & = & {1\over 2}
\left[A_0^2 f_0(r) - 3 A_r^2 f_r(r) \right]~~. \eeqa

Hence, one must solve this set of coupled second order
differential equations, with boundary conditions given by
$\phi(\infty) = \rho(\infty) = \phi'(\infty) = \rho'(\infty) = 0$.

Before continuing, we point out that LSB is clearly exhibited:
noticing that $g_0 + g_r = f_0 + f_r $, one has $ (1 - 2A_0^2) f_0
= - ( 1 + 2A_r^2) f_r$, which is an explicit manifestation of the
breaking of Lorentz symmetry. In the unperturbed case $A_0 = A_r =
0$, $f_0 = - f_r$ and one recovers the Schwarzschild solution
$\phi = -\rho$ from the equation $g_0 + g_r = 0 $. This symmetry
does not hold in the perturbed case, which yields $f_0 \approx
-(1+2A_0^2+2A_r^2)f_r$.

To solve the set of coupled differential equations Eqs.
(\ref{coupled}) one considers an expansion of the metric in terms
of $\phi=\phi_0 + \de \phi$ and $\rho = - \phi_0 - \de \rho$,
where $\phi_0$ is given by the usual Szcharzschild metric:

\beq \phi_0 = {1 \over 2} ln \left(1- {2M \over r} \right) ~~,
\eeq

\noindent and $\de \rho$, $\de \phi$ are assumed to be small
perturbations. Hence, one gets to first order

\beq f_0(r) = {2 \over r} \de \phi' + { {M \over r^2} \over 1 -
{2M \over r} } (3 \de \phi' + \de \rho') + \de \phi'', \eeq

\noindent and

\beq f_r(r) = -{2 \over r} \de \rho' - { {M \over r^2} \over 1 -
{2M \over r} } (3 \de \phi' + \de \rho') - \de \phi'' ~~. \eeq

\noindent \noindent As expected, the above quantities are
homogeneous in $\de \rho$, $\de \phi$ and their derivatives.

For the calculus of $g_0(r)$ and $g_r(r)$, one first computes the
contribution of the exponential term:

\beqa && {1 - e^{2 \rho} \over r^2} = {1 - e^{-2 (\phi_0 + \de
\rho)} \over r^2} \\ \nonumber & \simeq & {1 - e^{-2\phi_0} (1 - 2
\de \rho) \over r^2} = { {2M \over r^3} \over 1 - {2M \over r}} +
{2 \over 1 - {2M \over r} } {\de \rho \over r^2}~~. \eeqa

\noindent Thus, one finds

\beq g_0(r) = - {2 \over r} \left( {1 \over 1 - {2M \over r}} {\de
\rho \over r} + \de \rho' \right) \eeq

\noindent and

\beq g_r(r) =  {2 \over r} \left( {1 \over 1 - {2M \over r}} {\de
\rho \over r} + \de \phi' \right)~~, \eeq

\noindent which are also homogeneous.

To solve Eqs. (\ref{coupled}), one first obtains a relation
between $\de \rho$ and $\de \phi$. For this, one sums both coupled
differential equations, which yields

\beqa && {2 \over r} (\de \phi' - \de \rho') = 4 [ {2 \over r} (A
\de \phi' + B \de \rho') \\ \nonumber & + & (A + B) { {M \over
r^2} \over 1 - {2M \over r} } (3 \de \phi' + \de \rho')  +
(A+B)\de \phi'' ] ~~, \eeqa

\noindent where the simplifying notation $A=A_0^2$, $B=A_r^2$ has
been used ($A$ and $B$ are dimensionless). Since the LSB is
presumably a small effect, one expects $A \ll 1$ and $B \ll 1$.
Hence, one has

\beqa & & \de \rho' \left[ 1 + 4B + {A+B \over 2} {{M \over r}
\over 1 - {2M \over r}} \right] \\  \nonumber &=& \de \phi' \left[
1 - 4A - {3(A+B) \over 2} {{M \over r} \over 1 - {2M \over r}}
\right] - {A+B \over 2} r \de \phi''~~. \eeqa

\noindent Dropping terms smaller than $O(A)$, $O(B)$, this
expression simplifies to

\beq \de \rho'= { 1 - 4A \over 1 + 4B} \de
\phi' - {A+B \over 2(1 + 4B)} r \de \phi'' ~~, \eeq

\noindent which, after integration, yields

\beq \de \rho = {2 + B - 7A \over 2(1+4B)} \de \phi - {A+B \over 2
(1 + 4B)} r \de \phi' ~~. \eeq

One can now introduce this expression in one of the Eqs.
(\ref{coupled}) and obtain, after a somewhat tedious computation,
the following ordinary differential equation

\beq -C_1 r^2 \de \phi'' + C_2 r \de \phi' + C_3 \de \phi = 0~~,
\eeq

\noindent with

\beqa \nonumber C_1 & = & A + 3B +AB + 9 B^2 \simeq A + 3B ~~,
\\ \nonumber C_2 & = & 2+B-3A+16AB \simeq 2~~, \\ C_3 & = & 2+B-7A
\simeq 2 ~~. \eeqa

This equation has the solution

\beqa \nonumber \de \phi & = & K_- r^{C_1 + C_2 - \sqrt{(C_1 +
C_2)^2 + 4 C_1 C_3} \over 2 C_1} \\ & + & K_+ r^{C_1 + C_2 +
\sqrt{(C_1 + C_2)^2 + 4 C_1 C_3} \over 2 C_1}~~, \eeqa

\noindent where $K_-$ and $K_+$ are integration constants. Since
all $C_i$ are positive, one obtains that $\sqrt{(C_1 + C_2)^2 + 4
C_1 C_3} > C_1 + C_2 $, and hence the second power-law term
diverges when $r \rightarrow \infty$. Since one demands that $\de
\phi \rightarrow 0$, this implies that $K_+=0$. The remaining
first power-law term automatically satisfies the conditions $\de
\phi \rightarrow 0$ and $\de \phi' \rightarrow 0$ for all $K_-
\equiv K$. Hence, the perturbation is simply given by $\de \phi =
K r^{-\al}$, where

\beq \al = {-C_1 - C_2 + \sqrt{(C_1 + C_2)^2 + 4 C_1 C_3} \over 2
C_1} > 0~~, \eeq

\noindent and $K$ has dimensions of $[K] = L^\al$. One can
linearize this exponent with respect to $C_1 \ll 1$, that is

\beq \al \simeq {C_3 \over C_1 + C_2} = {2-7A+B \over 2+A+3B}~~,
\eeq

\noindent so that $\al \simeq 1$. One can now compute $\de \rho$,
obtaining

\beqa \nonumber \de \rho & = & {2 + B - 7A \over 2(1+4B)} \de \phi
- {A+B \over 2 (1 + 4B)} r \de \phi' \\ & \simeq & \left[ 1 +
\al{(A+B) \over 2} \right] K r^{-\al} ~~. \eeqa

\par Hence, the non-trivial components of the metric read

\beqa g_{tt} &=& -e^{2 (\phi_0 + \de \phi)} = -e^{2K r^{-\al}}
\left( 1 - {2M \over r} \right)~~, \\ \nonumber g_{rr} &=&
e^{-2(\phi_0 + \de \rho)} = {e^{-\left(2 + \al(A+B) \right) K
r^{-\al}} \over 1 - {2M \over r}} \equiv {e^{-2K_r r^{-\al}} \over
1 - {2M \over r}} ~~, \label{smetric} \eeqa

\noindent where one defines $K_r \equiv [1 + \al(A+B)/2]K \simeq
[1 + (A+B)/2]K$. To compute the PPN parameters, one first performs
a Lorentz transformation to a isotropic coordinate system, that
is, to one on which all spatial metric components are equal. Since
the angular coordinates are not affected by the LSB dynamics, this
amounts only to a change of the radial parameter, $r \rightarrow
\xi = \xi(r)$. Thus, instead of explicitly deriving the associated
metric $\bar{g}_{\mu\nu}$ through its transformation properties,
one resorts to the invariance of the interval $ ds^2 $, obtaining

\beqa \nonumber && g_{tt}(r) dt^2 + g_{rr}(r) dr^2 + r^2 d \Om^2 \\
&=& g_{tt}(\xi) dt^2 + \la(\xi) ( d \xi^2 + \xi^2 d\Om^2 ) ~~.
\eeqa

Equating the angular part, one obtains $\la(\xi) = r^2 / \xi^2$.
Since the temporal part is equal on both sides of the equation
(merely expressed in terms of $r$ or $\xi$), one also obtains $
g_{rr} dr^2 = \la(\xi) d\xi^2 $ and

\beq {g_{rr}^{1/2} \over r} dr = {d \xi \over \xi} ~~,
\eeq

\noindent implying that

\beq \int {e^{- K_r r^{-\al}} \over r \sqrt{1-{2M \over r}}}dr =
log~\xi + const~~. \eeq

\noindent Since the perturbation is assumed to be small, one can
consider that both $K, ~K_r \ll r^\al$. Hence, one expects $\xi$
also to be a small perturbation to the unperturbed isotropic
coordinate $\xi_0$; indeed, this is obtained setting $K_r=0$, from
which follows

\beq {dr \over \sqrt{r^2 - 2M r}} = {d \xi_0 \over \xi_0} ~~\eeq

\noindent and

\beqa \nonumber r & = & \xi_0 \left(1+{M \over 2 \xi_0}\right)^2 ~~, \\
\xi_0 & = & {1 \over 2} \left[ r \left( 1 + \sqrt{ 1 - {2M \over
r} } \right) - M \right]~~. \eeqa

Hence, to solve the perturbed $K_r \neq 0 $ case, one performs the
coordinate transformation $r \rightarrow \xi_0$, obtaining

\beq \int {e^{-K_r r^{-\al}} \over r \sqrt{1-{2M \over r}}}dr =
\int {e^{-K_r r^{-\al}(\xi_0)} \over \xi_0} d \xi_0 ~~.\eeq

Expanding the integrand around $\al = 1 $ (through $r^\al = r +
(\al-1) r ~log(r)$) and $ K_r = 0$, it can be easily seen that the
contribution of the $\al$ term amounts to corrections of
second-order $ O \left( (\al - 1) K_r \right) $, which we shall
disregard. This is equivalent to set $\al=1$ in the above
expressions. Hence, one obtains

\beqa \nonumber log~\xi & = & \int {e^{-K_r r^{-\al}(\xi_0)} \over
\xi_0} d \xi_0 \simeq \int {1 - {K_r \over r(\xi_0)} \over \xi_0}
d \xi_0 \\ & = & log(\xi_0) + {K_r \over \xi_0 + {M \over 2}}
~~.\eeqa

Solving for $\xi$, yields

\beq \xi = \xi_0 e^{K_r \over \xi_0 + {M \over 2}} ~~.\eeq

\noindent Obviously, setting $K_r = 0$ gives $\xi=\xi_0$, that is,
one recovers the Schzarschild isotropic coordinates. One can now
read the (isotropic) spatial component of the metric
$r^{-2}\bar{g}_{\th\th} = r^{-2}sin^{-2}(\th)
\bar{g}_{\phi\phi}=\bar{g}_{rr}$:

\beq \bar{g}_{\xi\xi} = \la (\xi) = e^{-{2 K_r \over \xi_0 + {M
\over 2}}} \left(1+{M \over 2 \xi_0}\right)^4 ~~. \eeq

\noindent Recall that the Lorentz transformation maps $r
\rightarrow \xi$, while $\xi_0$ is just a convenient integration
variable. Hence, one must now write $\xi_0$ in terms of $\xi$. For
this, one should invert the relation

\beq \xi = \xi_0 e^{K_r \over \xi_0 + {M \over 2}} \simeq \xi_0 +
{ K_r \over 1 + {M \over 2 \xi_0}}  ~~,\eeq

\noindent and hence, to first order in $K_r$,

\beq \xi_0 \simeq \xi - K_r \left( 1- {M \over 2 \xi} \right)~~.
\eeq

Inserting this expression into the above equation and expanding to
first order in $U = M / \xi$, one gets

\beqa \nonumber && \bar{g}_{\xi\xi} = e^{-{2 K_r \over \xi_0(\xi)
+ {M \over 2}}} \left(1+{M \over 2 \xi_0 (\xi) }\right)^4 \\
& \simeq & 1 + 2\left( 1 - {K_r \over M} \right) U ~~. \eeqa

Through a similar procedure, the $\bar{g}_{tt}$ component can be
found:

\beqa \nonumber \bar{g}_{tt} & = & -e^{2K r^{-\al}(\xi)} \left( 1
- {2M \over r(\xi)} \right) \\ & = & -e^{2K \xi_0^{-1} \left(1+{M
\over 2 \xi_0}\right)^{-2}} {\left(1 - {M \over 2 \xi_0}\right)^2
\over \left(1 + {M \over 2 \xi_0}\right)^2} \\ \nonumber & \simeq
& -1 + 2 \left( 1 - {K \over M} \right) U - 2 \left( 1 - {3K + K_r
\over M} \right) U^2 ~~, \eeqa

\noindent where, as before, we have taken $\al=1$ (so that
$K,~K_r$ have dimension of length). Both $\bar{g}_{tt}$ and
$\bar{g}_{\xi\xi}$ reduce to the usual isotropic metric components
in the limit $K,~K_r \rightarrow 0$.

To read the PPN parameters one must now transform to a
quasi-cartesian referential, where the metric reads

\beqa \nonumber \eta_{tt} & = & -1 +2U - 2 \be U^2~~, \\ \eta_{ij}
& = & (1 + 2 \ga U) \de_{ij}~~.\eeqa

\noindent This is achieved by a (spatial) coordinate change $\xi
\rightarrow \xi' $ so that $\xi = (1 - K/M) \xi'$. This transforms
the metric to

\beqa \nonumber \eta_{tt} & = & g_{tt} = -1 + 2 {M \over \xi'} - 2
{ 1 - {3K + K_r \over M}  \over \left(1-{K \over M}\right)^2 }
\left({ M \over \xi'} \right)^2 \\ \nonumber & \simeq & -1 + 2 U -
\left(1- {K + K_r \over M} \right) U^2~~,
\\ \nonumber \eta_{\xi' \xi'} & = & \left( \partial \xi \over \partial
\xi' \right)^2 \bar{g}_{\xi\xi} \\ \nonumber & = & \left( 1 - {K
\over M} \right)^2 \left[ 1 + 2 \left( 1 - {2 K_r \over M} \right)
{M \over (1 - K/M) \xi'} \right] \\ & \simeq & 1 + \left(1 - {K +
2 K_r \over M} \right) U ~~.\eeqa

Hence, one obtains

\beqa \nonumber \be & \simeq & 1 - {K + K_r \over M} ~~, \\
\ga & \simeq & 1 - {K + 2 K_r \over M} ~~. \eeqa

\noindent Inverting, one finds that

\beqa \nonumber {K \over M} \simeq 1 - 2 \be + \ga ~~, \\ {K_r
\over M } \simeq \be - \ga~~. \eeqa

\noindent A drawback of these results is the dependence of these
PPN parameters on $K$ and $K_r$, which are integration constant
(more precisely, $K_r$ is defined as proportional to $K$, which is
free valued). Therefore, these do not depend on the physical
parameters associated with the breaking of Lorentz invariance.
This reflects the linearization of the Einstein's equation that
was used in order to obtain the radially symmetric Birkhoff
metric. In any case, one can conclude that the effect of
temporal/radial LSB manifests itself linearly on the PPN
parameters $\be$ and $\ga$.

The current bounds, arising from the Nordvedt effect, $| \be -1 |
\leq 6 \times 10^{-4}$ \cite{Will2} and the Cassini-Huygens
experiment, $ \ga = 1 + (2.1 \pm 2.3) \times 10^{-5}$
\cite{Bertotti}, can be used to constraint the parameter space
$(K,K_r)$. Taking the Sun's geometrical mass $ M = 1.5~km$, one
obtains the constraints

\beqa \nonumber && |K + K_r| < 0.9~m ~~,\\ && K + 2 K_r = (-3.1
\pm 3.4) \times 10^{-2}~m~~. \eeqa

\begin{figure}

\epsfysize=5.8cm \epsffile{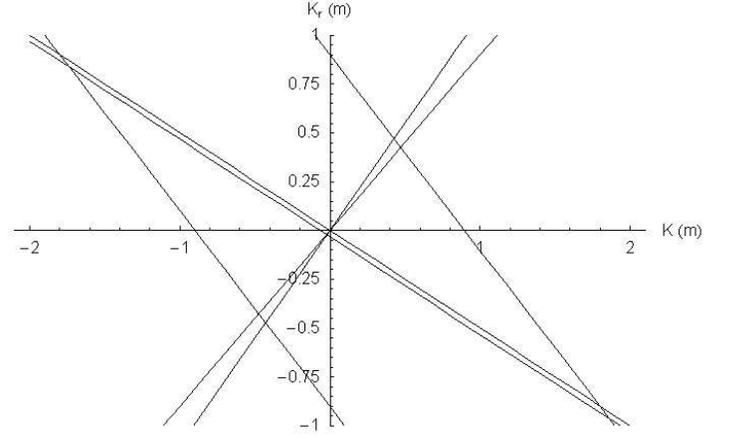} \caption{Allowed values for
$K$ and $K_r$, derived from $| \be -1 | \leq 6 \times 10^{-4}$, $
\ga = 1 + (2.1 \pm 2.3) \times 10^{-5}$ and the assumption $|1 -
K_r / K | \lesssim 0.1$.} \label{fig1}

\end{figure}

\begin{figure}

\epsfysize=5.8cm \epsffile{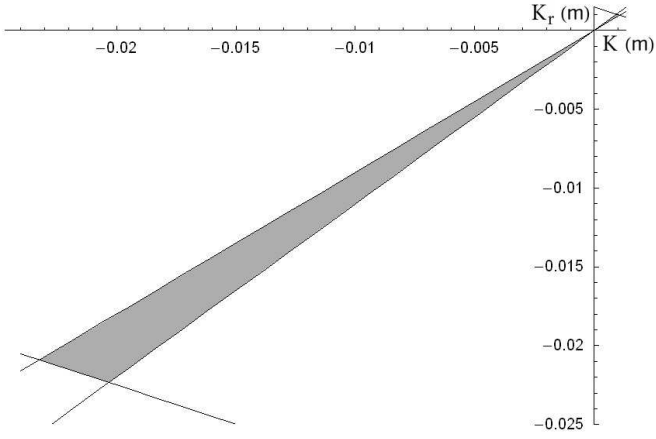} \caption{Detail of Fig.
\ref{fig1}, showing only the allowed region.} \label{fig2}

\end{figure}

\noindent Also, by definition

\beq K_r = \left[ 1 + \al{(A+B) \over 2} \right] K~~, \eeq

\noindent with $\al \simeq 1$, $A , B \ll 1$. Hence, one expects
deviations of $K_r$ from $K$ to be small. Thus, considering for
instance the constraint $|1 - K_r / K | \lesssim 0.1$, the
resulting range of allowed values for these parameters is depicted
in Figures \ref{fig1} and \ref{fig2}. Notice that considering $K
\sim K_r$ immediately yields $K \simeq (-1 \pm 1.1) \times
10^{-2}$, indicating that the perturbation has a very short range.
Indeed, since this lies well inside the Sun, one should work with
the interior Scharzschild solution, and this value merely
indicates that the effect outside the Sun is negligible.

Returning to Eqs. (\ref{smetric}) one notices that, in the limit
$M \rightarrow 0$, one gets $\bar{g}_{tt} = -e^{2K / \xi}$ and
$\bar{g}_{rr} = -e^{2K_r / \xi}$. This allows establishing an
analogy with Rosen's bimetric theory, by interpreting these
changes as due to a background metric $\eta_{\mu\nu}$
\cite{Rosen}. Notice that, in the absence of a central mass, the
vector field no longer rolls to a radial \textit{vev}, since this
spatial symmetry is inherited from the vanishing covariant
derivative, that is, from the presence of the central mass. With
this in mind, one can pursue the analogy with Rosen's bimetric
theory and read the PPN parameter $\al_2$ \cite{Will}:

\beq \eta_{\mu\nu} = diag\left(
-c_0^{-1},c_1^{-1},c_1^{-1},c_1^{-1} \right) \equiv
\bar{g}_{\mu\nu~(M\rightarrow 0)} ~~, \eeq

\noindent so that

\beqa \nonumber \al_2 & = & {c_0 \over c_1} -1 = e^{2(K_r - K) / \xi} - 1 \\
& \simeq & { 2 (K_r - K) \over \xi} = (A + B) {\al \over \xi }
~~.\eeqa

\noindent which exhibits a radial dependency. Assuming $\al \simeq
1$ and considering the spin precession constraint arising from
solar to ecliptic alignment measurements \cite{Will2}, one has $|
\al_2 | = |A+B|/r < 4 \times 10^{-7}$, which implies $|A+B| < 4
\times 10^{-7} r_\odot = 2.78 \times 10^2~m $, where $r_\odot =
6.96 \times 10^{8}~m$ is the radius of the Sun. This also reflects
the smallness of the expected effect on the metric.

Since there is no explicit Lorentz breaking, the speed of light
remains equal to $c$. The speed of gravitational waves, on the
other hand, is shifted by an amount

\beq \sqrt{c_1 \over c_0} = e^{2(K-K_r) / \xi} \simeq \left[ 1 -
{A + B \over \xi} \right]~~,\eeq

\noindent and hence acquires a radial dependence.

Let us close remarking that a direct comparison of the results of
this section with the ones of the purely radial case is
unfeasible. Indeed, the radial LSB effects are exact, while the
radial/temporal results are not. Therefore, one cannot directly
consider the $A \rightarrow 0$ limit for comparison of the results
obtained.

\section{Axial/temporal LSB}

One now treat the anisotropic LSB case. For definitiveness, one
assumes that the bumblebee field is stabilized at its vacuum
expectation value and exhibits both a temporal and a spatial
coordinate, assumed to lie on the x-axis, that is, $ b_\mu =
\ka^{-1} (a,b,0,0) $. One calculates the metric perturbations
$h_\mn$ to the flat Minkowsky metric. To first order in $h_\mn$,
one has

\beqa R_{00} & = & - {1 \over 2} \nabla^2 h_{00}~~, \\
\nonumber R_{0i} & = & -{1 \over 2} \left( \nabla^2 h_{0i} - h_{k0,ik} \right) ~~,
\\ \nonumber R_{ij} & = & -{1 \over 2} \left( \nabla ^2 h_{ij} - h_{00,ij} +
h_{kk,ij} - h_{ki,kj} - h_{kj,ki} \right) ~~, \eeqa

\noindent where time derivatives were neglected, since one assumes
that $v \ll c$.

The axial LSB breaks the radial symmetry, and one must consider
the Einstein equations: $ G_\mn = T_\mn + T_{B \mn}$, where $T_{B
\mn}$ is the stress-energy tensor for the bumblebee field,

\beq T_{B \mn} = \left[ {1 \over 2} b^\al b^\be R_{\al \be} g_\mn
- b_\mu b^\al R_{\al \nu} - b_\nu b^\al R_{\al \mu} \right]~~.\eeq

\noindent Its trace clearly vanishes and hence, from the trace of
the Einstein equations one obtains

\beq R_\mn = \ka \left[T_\mn + {1 \over 2} g_\mn T + T_{B \mn}
\right] ~~. \eeq

First one calculates $h_{00}$ to first order of the potential $U$;
for that one writes

\beq R_{00} =  -{1 \over 2} \left( a^2 R_{00} + b^2 R_{11} + 2 a b
R_{10} \right) - 2 a \left( a R_{00} + b R_{10} \right) ~~. \eeq

\noindent Since

\beq R_{00} = - {1 \over 2} \nabla^2 h_{00}~~,~~ R_{0i} = 0~~,~~
R_{ij} = {1 \over 2}h_{00,ij} ~~, \eeq

\noindent one has

\beq R_{00} = -{1 \over 2} \left( a^2 R_{00} + b^2 R_{11} \right)
- 2 a^2 R_{00} ~~, \eeq

\noindent and hence

\beq \left( 1 + {5 a^2 \over 2} \right) \nabla^2 h_{00} - {b^2
\over 2} h_{00,11} = 0~~. \eeq

\noindent One can rewrite this as

\beq \left({2 + 5 a^2 - b^2 \over 2 + 5 a^2} \right) h_{00,11} +
h_{00,22} + h_{00,33} = 0 ~~. \eeq

\noindent This equation admits the solution

\beq h_{00} (x,y,z) = {2 M \over \sqrt{c_0^2 x^2 + y^2 + z^2}} ~~,
\eeq

\noindent where $c_0^2 = ( 2 + 5 a^2) / (2 +5 a^2 - b^2)$.

One now computes the components $h_{ii}$ ($i \neq 1$) using this
result:

\beq R_{ii} = {1 \over 2} \left(a^2 R_{00} + b^2 R_{11}  + 2 a b
R_{01} \right) ~~, \eeq

\noindent hence

\beqa \nonumber R_{00} & = & - {1 \over 2} \nabla^2 h_{00}~~, \\
\nonumber R_{01} & = & 0 ~~, \\ \nonumber  R_{ii} & & -{1 \over 2}
\left(\nabla ^2 h_{ii} - h_{00,ii} + h_{ii,ii} \right) ~~, \\
R_{11} & = & {1 \over 2} h_{00,11}~~, \eeqa

\noindent and therefore (for $j \neq i$, $ j \neq 1$)

\beqa & & -{1 \over 2} \left( \nabla ^2 h_{ii} - h_{00,ii} +
h_{ii,ii} \right) \\ \nonumber & = & {a^2 \over 2} \left( - {1
\over 2} \nabla^2 h_{00} \right) + {b^2 \over 2} \left({1 \over 2}
h_{00,11} - {1 \over 2} h_{ii,11} \right)~~. \eeqa

\noindent One rewrites this equation as

\beqa & & \left( 2 - b^2 \right) h_{ii,11} + 4 h_{ii,ii} + 2
h_{ii,jj} \\ \nonumber &=& \left( a^2 - b^2 \right) h_{00,11} +
\left( 2 + a^2 \right) h_{00,ii} + a^2 h_{00,jj} ~~.
\label{partial} \eeqa

\noindent As an Ansatz for the solution, one takes $h_{ii} (x,y,z)
= h_{00} (\al_1 x, \al_2 y, \al_3 z)$. Notice that

\beq h_{ii,jj} = \al_j^2 h_{00,jj}(\al_1 x, \al_2 y, \al_3 z)
\simeq \al_j^2 h_{00,jj}(x, y, z) \label{ansatz} \eeq

\noindent and that

\beqa \nonumber h_{00,11} & = & - 2 M c_0^2 { - 2 c_0^2 x^2 + y^2
+ z^2 \over \left({c_0^2 x^2 + y^2 + z^2} \right)^5 }~~, \\
h_{00,jj} & = & - 2 M { - 2 x_j^2 + c_0^2x^2 + x_k^2 \over
\left({c_0^2 x^2 + y^2 + z^2} \right)^5 } ~~. \label{second} \eeqa

\noindent Substituting in Eq. (\ref{ansatz}) one obtains, after
some calculation (see Appendix I), the coefficients $\al_i$:

\beqa \nonumber \al_1^2 & = & {-2 ( 2 + 5a^2 - b^2) \al_j^2 + 2 (1
+ 2a^2)b^2 \over (2 + 5 a^2 ) ( -2 + b^2)}~~, \\ \al_i^2 & = & {1
+ \al_j^2 \over 2}~~, \eeqa

\noindent for free $\al_j$. In order to match with the unperturbed
case $r^2 = x_i x^i$ one chooses $\al_j=1$ to get

\beq \al_1^2 = 1 - {(2 - a^2) b^2 \over (2 + 5 a^2) ( 2 - b^2 )
}~~, ~~ \al_i = \al_j = 1 ~~. \eeq

\noindent Hence,

\beq h_{ii} (x,y,z) = h_{00} (\al_1 x, y, z) \equiv {2 M \over
\sqrt{c_2^2 x^2 + y^2 + z^2}} ~~, \eeq

\noindent where it has been defined

\beq c_2^2 = \al_1^2 c_0^2 = {2 (2 + 5a^2 - 2 b^2) - 4 a^2b^2
\over (2 + 5a^2 - b^2) (2 - b^2)} ~~. \eeq

We now compute the $h_{11}$ component:

\beq R_{11} = {1 \over 2} \left( a^2 R_{00} + b^2 R_{11} \right) -
2b^2 R_{11}~~, \eeq

\noindent leading to

\beq  \left( 1 + {3 b^2 \over 2} \right) R_{11} = {a^2 \over
2}R_{00}~~, \eeq

\noindent and

\beqa R_{00} & = & -{1 \over 2} \nabla^2 h_{00}~~ , \\
\nonumber R_{11} & = & - {1 \over 2} \left( \nabla^2 h_{11} -
h_{00,11} + h_{kk,11} - 2 h_{11,11} \right) ~~. \eeqa

\noindent As before, one can combine these equations into

\beqa \nonumber && h_{11,22} + h_{11,33} = {2 + a^2 + 3 b^2 \over
2 + 3 b^2} h_{00,11} \\ & + & {a^2 \over 2 + 3 b^2} (h_{00,22} +
h_{00,33}) - 2 h_{22,11}~~. \eeqa

\noindent Introducing the expression for $h_{00}$ and $h_{22}$ in
the r.h.s. term, one gets

\begin{widetext}

\beqa && h_{11,22} + h_{11,33} = 2M \left[ \left({ a^2 (c_0^2 -1)
\over 2 + 3 b^2 } + c_0^2 \right) { 2 c_0^2 x^2 -y^2 -z^2 \over
(c_0^2 x^2 + y^2 + z^2 )^{5/2} }- 2 c_2^2 {2 c_2^2 x^2 -y^2 -z^2
\over (c_2^2 x^2 + y^2 + z^2 )^{5/2} } \right]~~. \eeqa

\end{widetext}

\noindent It is clear that the solution is a linear combination of
$h_{00}$ and $h_{22}$. Indeed, one has

\beqa & & h_{11} (x,y,z) \\ \nonumber & = & - \left({ a^2 ( c_0^2
- 1 ) \over 2 + 3 b^2 } + c_0^2 \right) h_{00}(x,y,z) + 2 c_2^2
h_{22} (x,y,z)~~. \eeqa

One now searches for the off-diagonal component $h_{10}$. The
component

\beq R_{10} = -(a^2 + b^2) R_{10} - ab (R_{00} + R_{11}) ~~, \eeq

\noindent leads to

\beq (1 + a^2 + b^2) R_{10} = -ab (R_{00} + R_{11}) ~~, \eeq

\noindent and

\beqa R_{00} & = & - {1 \over 2} \nabla^2 h_{00}~~, \\
\nonumber R_{10} & = & - {1 \over 2} (\nabla^2 h_{10} - h_{10,11}
) ~~, \\ \nonumber R_{11} & = & - {1 \over 2} (\nabla^2 h_{11} -
h_{00,11} + h_{kk,11} - 2 h_{11,11} )~~. \eeqa

\noindent Repeating the above procedure, one obtains

\beqa  && h_{01,22} + h_{01,33} = -{ab \over 1+ a^2 + b^2} \\
\nonumber & \times & \left[ h_{00,22} + h_{00,33} + h_{11,22} +
h_{11,33} - 2 h_{22,11} \right] ~~. \eeqa

\noindent Writing $h_{01} = -ab/(1+a^2+b^2)(h_{00} + h_{11} + \de
h_{01})$, one finds

\beqa \nonumber & & \de h_{01,22} + \de h_{01,33} = 2 h_{22,11} \\
& = & {4 M c_2^2 (2 c_2^2 x^2 - y^2 - z^2) \over (c_2^2 x^2 +y^2
+z^2)^{5/2}}~~, \eeqa

\noindent and hence

\beq \de h_{01}(x,y,z) = - 2 c_2^2 h_{22} (x,y,z)~~. \eeq

\noindent Therefore,

\beq h_{01} = {a b \over 1 + a^2 + b^2 } \left[{ a^2 ( c_0^2 - 1 )
\over 2 + 3 b^2 } + c_0^2 \right] h_{00} ~~ . \eeq

Finally, one computes $h_{00}$ to second order. The results are
developed in the Appendix II. It is found that one gets only a
correction to the first order term $h_{00}^{(1)}$:

\beqa \nonumber h_{00} & = & {2c_0^2 [6 + 9 b^2 + (15 + 22 b^2)
a^2] + a^2 b^2 \over c_0^2 (6 + 15 a^2 + b^2) (2 +3b^2)} h_{00}^{(1)} \\
& \simeq & \left( 1 - {b^2 \over 6} \right) h_{00}^{(1)}~~.\eeqa

Since the LSB clearly turns the metric anisotropic, the usual PPN
parametrization cannot be straightforwardly used to ascertain its
effects. This is so as the PPN formalism relies on a
quasi-cartesian frame of reference which, by definition, has all
diagonal metric components $g_{ii}$ equal. One might argue that
there is a transformation to such a isotropic frame, but the
obtained PPN parameters would undoubtedly be unphysical. However,
one can extract some PPN-like parameters from the results. First,
one notes that

\beq {1 \over \sqrt{(1+ 2 k) x^2 + y^2 + z^2}} \simeq {1 \over r}
\left( 1 - k cos^2 \th \right) ~~ . \eeq

For $h_{00}$, one gets

\beq h_{00} = \left( 1 - {b^2 \over 6} \right) 2M {1 - (c_0^2 - 1)
cos^2 \th \over r}~~, \label{anisotropy} \eeq

\noindent where no $r^{-2}$ correction appears. Thus, one cannot
obtain the PPN parameter $\be$. However, since $h_{11} \neq h_{22}
= h_{33}$, the same reasoning allows us to compute two parameters
analogous to the $\ga$ PPN parameter. Recalling that

\beqa \nonumber &&  h_{11} (x,y,z) = - \left[{ a^2 ( c_0^2 - 1 )
\over 2 + 3 b^2 } + c_0^2 \right] h_{00}(x,y,z) \\ & + & 2 c_2^2
h_{22} (x,y,z)~~, \eeqa

\noindent one gets (after neglecting the normalization with
respect to $h_{00}$),

\beqa \nonumber \ga_1 & = & 1 + cos^2 \th \\ \nonumber & \times &
\left[ -\left[{ a^2 ( c_0^2 - 1 ) \over 2 + 3 b^2 } + c_0^2
\right] (1 - c_0^2) + 2 c_2^2 (1 -
c_2^2) \right] \\ \nonumber & \simeq & 1 + {b^2 \over 2} cos^2 \th ~~, \\
\ga_2 &=& 1 + (1 - c_2^2) cos^2 \th \simeq 1 - \left({a b \over 2}
\right)^2 cos^2 \th~~. \eeqa

As expected, the effect of the x-axis LSB is mostly felt on the
$h_{11}$ component. Direct comparison with the PPN parameter $\ga$
is troublesome, given that the present case is obviously
anisotropic. Hence, no clear connection can be derived to link
$\ga$ with $\ga_1$ and $\ga_2$.

However, the measured value of $\ga$ should be of the same order
of magnitude as the average of $\ga_1$ and $\ga_2$, integrated
over one orbit:

\beq \ga -1 \simeq {1 \over 2} (\ga_1 + \ga_2) - 1 \simeq  {b^2
\over 4} \langle cos^2 \th \rangle \simeq {b^2 \over 8}
(1-e^2)~~,\eeq

\noindent where $e$ is the orbit eccentricity. For low values of
$e$ such as those found in the Solar System, one gets $\ga \simeq
b^2 / 8$. Taking $e \simeq 0$, the constraint $\ga =1 + (2.1 \pm
2.3) \times 10^{-5}$ now yields $|b| \leq 1.9 \times 10^{-2}$.

As stated above, the standard PPN analysis fails in the present
scenario, which is clearly anisotropic. A discussion involving
anisotropy of inertia and its effect in the width of resonance
lines can be found in Ref. \cite{Weinberg} and references therein
(see also Ref. \cite{Kostelecky6}). Presented as a test between
Mach's principle and the equivalence principle, it relies on the
hypothetical effect the proximity to the large mass of the
galactic core could have on the proton's mass. By comparison, we
note that a radial LSB with the galactic core as source would be
perceived as an axial LSB in a region such as the Solar System. In
Ref. \cite{Lamoreaux} the bound $\De m / m \leq 3 \times 10^{-22}$
can be found, $m$ being the proton mass. Comparing with Eq.
(\ref{anisotropy}) gives

\beq {\De m \over m} = \left( 1 - {b^2 \over 6} \right) (c_0^2 -1
) \simeq {b^2 \over 2} \leq 3 \times 10^{-22}~~,\eeq

\noindent which yields $ |b| \leq 2.4 \times 10^{-11} $, a much
more stringent bound than the obtained above.

\section{Conclusions}

In this study we have obtained the solutions of a gravity model
coupled to a vector field where Lorentz symmetry is spontaneously
broken. We have analyzed three cases: purely radial,
temporal/radial and temporal/axial LSB.

In the first case, we have found a new black hole solution; we
showed that, as in the standard scenario, this solution has a
removable singularity at a horizon of radius $r_s = (2M
r_0^{-L})^{1/1-L}$, which is slightly perturbed with respect to
the usual Scharzschild radius $r_{s0} = 2M$. This horizon has an
associated Hawking temperature of $T = (2M r_0^{-L})^{-L}T_0$, and
protects a real singularity at $r=0$. Deviations from Kepler's law
yield $L \leq 2 \times 10^{-9}$.

The temporal/radial case yields a slightly perturbed metric which
leads to PPN parameters $ \be \approx 1 - (K + K_r)/ M$ and $\ga
\approx 1 - (K + 2 K_r) / M$, directly proportional to the
strength of the induced effect (given by $K$ and $K_r \sim K$).
Unfortunately, since $K$ and $K_r$ are integration constants, one
cannot constrain the physical parameters from the observed limits
on the PPN parameters. Also, analogously to Rosen's bimetric
theory, one can obtain the PPN parameter $\ga \simeq (A + B)\xi$,
where $\xi$ is the distance to the central body and $A$ and $B$
rule the temporal and radial components of the vector field
\textit{vev}.

In the temporal/axial case one gets, as expected, a breakdown of
isotropy, and hence a standard PPN analysis is not feasible.
However, we have defined the direction-dependent ``PPN''
parameters $ \ga_1 \simeq b^2 cos^2 \th /2$ and $\ga_2 \simeq a^2
b^2 cos^2 \th / 4$ (considering the LSB occurs in the direction of
$x_1$). Naturally, $\ga_1 \ll \ga_2$. A crude estimative of the
measurable PPN parameter $\ga$ yields $ \ga \approx b^2(1-e^2)/4$,
where $e$ is the orbit's eccentricity. This final case requires
further study, as its effects cannot be properly accounted for an
isotropic formalism such as the parametrized post-Newtonian one.
However, comparison with experiments concerning the anisotropy of
inertia yields $ |b| \leq 2.4 \times 10^{-11} $.


\begin{acknowledgments}

\noindent JP is sponsored by the Funda\c{c}\~{a}o para a
Ci\^{e}ncia e Tecnologia (Portuguese Agency) under the grant
BD~6207/2001.

\end{acknowledgments}

\section{Appendix I}

Substituting Eqs. (\ref{ansatz}) in Eq. (\ref{partial}), one gets
the following relation:

\begin{widetext}

\beq \left( (2 - b^2) \al_1^2 + (b^2 - a^2) \right) h_{00,11} +
\left( 4 \al_i^2 - (2 + a^2) \right) h_{00,ii} + \left( 2 \al_j^2
- a^2 \right) h_{00,jj} = 0 ~~, \eeq

\noindent and hence

\beqa \nonumber && \left( (2 - b^2) \al_1^2 + (b^2 - a^2) \right)
c_0^2 \left( - 2 c_0^2 x^2 + y^2 + z^2 \right) + \left( 4 \al_i^2
- (2 + a^2) \right) \left( - 2 x_i^2 + c_0^2x^2 + x_k^2 \right)  \\
& + & \left( 2 \al_j^2 - a^2 \right) \left( - 2 x_j^2 + c_0^2x^2 +
x_M^2 \right) = 0 ~~. \eeqa

For this equality to hold, the coefficients of each coordinate
must vanish, leading to

\beqa \nonumber & & 2(2 + 5 a^2) [-(2+5 a^2) ( 1 + 2 \al_1^2 - 2
\al_i^2 - \al_j^2) + (-1 + 2 \al_1^2 + a^2 (-4 + 5 \al_1^2) - 2
\al_i^2 - \al_j^2 ) b^2] = 0~~, \\ \nonumber & & - 2 ( 2 + 5 a^2)
(2 + \al_1^2 - 4 \al_i^2 + \al_j^2) + [a^2 (-4 + 5 \al_1^2) + 2 (1
+ \al_1^2 - 4 \al_i^2 + \al_j^2)] b^2 = 0 ~~, \\ & & -2 (2+ 5 a^2)
(-1 + \al_1^2 + 2 \al_i^2 - 2 \al_j^2) + [a^2 (-4 + 5 \al_1^2) + 2
(-2 + \al_1^2 + 2 \al_j^2 - 2 \al_3^2) ] b^2 = 0~~. \eeqa

These equations admit the solution

\beq \al_1^2 = {-2 ( 2 + 5a^2 - b^2) \al_j^2 + 2 (1 + 2a^2)b^2
\over (2 + 5 a^2 ) ( -2 + b^2)}~~, ~~ \al_i^2 = {1 + \al_j^2 \over
2}~~, \eeq

\noindent for a free $\al_j$.

\end{widetext}

\section{Appendix II}

To obtain $h_{00}$ to second order, one writes $h_{00} =
h_{00}^{(1)} + h_{00}^{(2)}$ and $R_{00} = R_{00}^{(1)} +
R_{00}^{(2)}$, $R_{ij} = R_{ij}^{(1)} + R_{ij}^{(2)}$. Notice that
$R_{0i} = R_{0i}^{(1)}$, since it contains no $h_{00}$ term. The
equation for $R_{00}^{(1)}$ is

\begin{widetext}

\beq R_{00}^{(1)} = -{1 \over 2} \left( a^2 R_{00}^{(1)} + b^2
R_{11}^{(1)} + 2 a b R_{10} \right) - 2 a^2 R_{00}^{(1)} - 2 a b
R_{10}~~. \eeq

\noindent One obtains

\beqa \nonumber & & R_{00}^{(1)} + R_{00}^{(2)} = -{1 \over 2}
\left( a^2 R_{00}^{(1)} + b^2 R_{11}^{(1)} + 2 a b R_{10} \right)
- 2 a^2 R_{00}^{(1)} - 2 a b R_{10} \\
\nonumber  & + &  {1 \over 2} \left( a^2 R_{00}^{(1)} + b^2
R_{11}^{(1)} + 2 a b R_{10} \right)h_{00}^{(1)} + {1 \over 2}
\left( a^2 R_{00}^{(2)} + b^2 R_{11}^{(2)}  \right) \left( -1 +
h_{00}^{(1)} \right) - 2 a^2 R_{00}^{(2)} \\ & = & \left[ ( 1 +
2a^2) R_{00}^{(1)} + 2 a b R_{10} \right] h_{00}^{(1)} + a^2
\left( - {5 \over 2} + h_{00}^{(1)} \right) R_{00}^{(2)} + {b^2
\over 2} \left( -1 + h_{00}^{(1)} \right) R_{11}^{(1)} ~~. \eeqa

\noindent Hence

\beq \left[ 1 + a^2 \left({5 \over 2} - h_{00}^{(1)} \right)
\right] R_{00}^{(2)} = \left( (1 + 2 a^2 ) R_{00}^{(1)} + 2 a b
R_{10} \right] h_{00}^{(1)} + {b^2 \over 2} R_{11}^{(2)} \left( -1
+ h_{00}^{(1)} \right) ~~, \eeq

\noindent and

\beqa \nonumber && \left[ 1 + a^2 \left({5 \over 2} - h_{00}^{(1)}
\right) \right] \left( - {1 \over 2} \nabla^2 h_{00}^{(2)} \right) \\
\nonumber & = & (1 + 2 a^2 ) \left[- {1 \over 2} \nabla^2
h_{00}^{(1)} + 2 a b \left( - {1 \over 2} \left( \nabla^2 h_{10} -
h_{10,11} \right) \right) \right] h_{00}^{(1)} \\ & + & {b^2 \over
2} \left[ - {1 \over 2} \left( \nabla^2 h_{11} - h_{00,11}^{(2)}
+2 h_{22,11} - h_{11,11} \right) \right] \left( -1 + h_{00}^{(1)}
\right) ~~. \eeqa

Dropping the $(-1 + h_{00}^{(1)})$ term yields

\beq \left( 1 + {5 a^2 \over 2} \right) \nabla^2 h_{00}^{(2)} =
-{b^2 \over 2} \left[ - {1 \over 2} \left( \nabla^2 h_{11} -
h_{00,11}^{(2)} +2 h_{22,11} - h_{11,11} \right) \right]~~. \eeq

\noindent It follows that

\beq \left( 1+ {5 a^2 + b^2\over 2} \right) h_{00,11}^{(2)} +
\left( 1+ {5 a^2 } \right) \left( h_{00,22}^{(2)} +
h_{00,33}^{(2)} \right) = - { b^2 \over 2} \left( h_{11,22}+
h_{11,33} + 2 h_{22,11} \right) ~~. \eeq

\noindent Taking the Ansatz $h_{00}^{(2)} = A h_{00}^{(1)}$, one
finds

\beq {2 M A c_0^2 (6 + 15 a^2 + b^2) \over 2 + 5 a^2} {2 c_0^2 x^2
- y^2 - z^2 \over \left( c_0^2 x^2 + y^2 + z^2 \right)^{(5/2)}} =
- 2M{ b^2 [a^2 (c_0^2 -1 ) + c_0^2 (2 + 3 b^2)] \over (2 + 5 a^2)
(2 +3b^2) } {2 c_0^2 x^2 - y^2 -z^2 \over (c_0^2 x^2 + y^2
+z^2)^{5/2}} ~~, \eeq

\noindent and

\beq A = - { b^2 [a^2 (c_0^2 -1 ) + c_0^2 (2 + 3 b^2)] \over c_0^2
(6 + 15 a^2 + b^2) (2 +3b^2) } ~~. \eeq

\noindent Thus, one obtains a small correction to $h_{00}^{(1)}$,
but no change in behavior. One can write

\beq h_{00} =  h_{00}^{(1)} + h_{00}^{(2)} = {2c_0^2 [6 + 9 b^2 +
(15 + 22 b^2) a^2] + a^2 b^2 \over c_0^2 (6 + 15 a^2 + b^2) (2
+3b^2)} h_{00}^{(1)} \simeq \left( 1 - {b^2 \over 6} \right)
h_{00}^{(1)}~~.\eeq

\end{widetext}


\begin{thebibliography}{99}


\bibitem{Kostelecky1} ``CPT and Lorentz Symmetry II'', V. A. Kosteleck\'{y} (World
Scientific, Singapore, 2002).

\bibitem{Bertolami} O. Bertolami, Nucl.\ Phys.\ Proc.\ Suppl.\ {\bf 88}, 49
(2000); O. Bertolami in ``Decoherence and Entropy in Complex
Systems'' (Springler-Verlag, Berlin, 2004).

\bibitem{Kostelecky2} V. A. Kosteleck\'{y} and S. Samuel, \PR {\bf D 39}, 683
(1989); \PRL {\bf 66}, 1811 (1991).

\bibitem{Kostelecky3} V. A. Kosteleck\'{y} and R. Potting, \PR {\bf D 51}, 3923 (1995).

\bibitem{Kostelecky4} V. A. Kosteleck\'{y}, \PR {\bf D 69}, 105009 (2004).

\bibitem{LSB1} R. Gambini, J. Pullin, \PR {\bf D 59}, 124021 (1999).

\bibitem{LSB2} J. Alfaro, H.A. Morales-Tecotl, L.F. Urrutia, \PRL
{\bf 84}, 2183 (2000).

\bibitem{LSB3} L.J. Garay, \PRL {\bf 80}, 2508 (1998).

\bibitem{LSB4} S.M. Carroll, J.A. Harvey, V.A. Kosteleck\'{y}, C.D. Lane,
T.Okamoto, \PRL {\bf 87}, 141601 (2001).

\bibitem{LSB5} O. Bertolami, L. Guisado, \PR {\bf D 67}, 025001 (2003).

\bibitem{LSB6} V.A. Kosteleck\'{y}, R. Lehnert, M. J. Perry, \PR {\bf D 68}, 123511
(2003) ; O. Bertolami, R. Lehnert, R. Potting, A. Ribeiro, \PR {
\bf D 69}, 083513 (2004).

\bibitem{LSB7} H. Sato, T. Tati, \PTP {\bf 47}, 1788 (1972); S.
Coleman, S.L. Glashow, \PL {\bf B 405}, 249 (1997); \PR {\bf D
59}, 116008 (1999); L. Gonzales-Mestres, hep-ph/9905430; O.
Bertolami, C.S. Carvalho, \PR {\bf D 61}, 103002 (2000); G.
Amelino-Camelia, T. Piran, \PL {\bf B 497}, 265 (2001); O.
Bertolami, \GRG {\bf 34} 707 (2002); R. Lehnert, hep-ph/0312093.

\bibitem{Kostelecky5} R. Bluhm and V. A. Kosteleck\'{y}, hep-th/0412320.

\bibitem{Rosen} N. Rosen, \textit{J. Gen. Rel. and Grav.} {\bf 4}, 435 (1973).

\bibitem{Will} C.M. Will, ``Theory and Experiment in Gravitational Physics'',
C.M. Will (Cambridge U. P., 1993).

\bibitem{Bento} M. C. Bento and O. Bertolami, \PL {\bf B 228}, 348
(1999).

\bibitem{Fischbach} ``The search for non-Newtonian gravity'', E.
Fischbach, C.L. Talmadge (Springer, New York 1999).

\bibitem{Will2} C. M. Will, Living Rev.\ Rel.\ {\bf 4}, 4 (2001).

\bibitem{Bertotti} B. Bertotti, L. Iess and P. Tortora, Nature {\bf 425}, 374 (2003).

\bibitem{Weinberg} S. Weinberg, ``Gravitation and Cosmology: Principles and Applications of the
General Theory of Relativity'' (John Wiley and Sons, New Jersey,
1972).

\bibitem{Kostelecky6} V. A. Kostelecky and C. D. Lane, J. Math. Phys. {\bf 40}
6245  (1999).

\bibitem{Lamoreaux} S. K. Lamoreaux, J. P. Jacobs, B. R. Heckel, F. J. Raab, and E. N. Fortson,
\PRL {\bf 58}, 746 (1987).


\end{thebibliography}
\end{document}